\documentclass[aps,prb,twocolumn,superscriptaddress]{revtex4}

\usepackage[dvips]{graphicx}
\usepackage{color}
\usepackage{threeparttable}
\usepackage{comment}

\begin{document}

\title{
Ferroelectricity of Dion-Jacobson layered perovskites CsNdNb$_2$O$_7$ and RbNdNb$_2$O$_7$
}

\author{Sota Asaki}
\affiliation{Department of Applied Chemistry, Kyushu University, Motooka, Fukuoka 819-0395, Japan}
\author{Hirofumi Akamatsu}
\email[e-mail:]{h.akamatsu@cstf.kyushu-u.ac.jp}
\affiliation{Department of Applied Chemistry, Kyushu University, Motooka, Fukuoka 819-0395, Japan}
\author{George Hasegawa}
\affiliation{Department of Applied Chemistry, Kyushu University, Motooka, Fukuoka 819-0395, Japan}
\author{Tomohiro Abe}
\affiliation{Graduate School of Science, Hiroshima University, Higashihiroshima, Hiroshima 739-8526, Japan}
\author{Yuki Nakahira}
\affiliation{Graduate School of Science, Hiroshima University, Higashihiroshima, Hiroshima 739-8526, Japan}
\author{Suguru Yoshida}
\affiliation{Department of Applied Chemistry, Kyushu University, Motooka, Fukuoka 819-0395, Japan}
\author{Chikako Moriyoshi}
\affiliation{Graduate School of Advanced Science and Engineering, Hiroshima University, Higashihiroshima, Hiroshima 739-8526, Japan}
\author{Katsuro Hayashi}
\email[e-mail:]{k.hayashi@cstf.kyushu-u.ac.jp}
\affiliation{Department of Applied Chemistry, Kyushu University, Motooka, Fukuoka 819-0395, Japan}

\date{\today}

\begin{abstract}

Crystallography and dielectric properties in Dion-Jacobson layered  perovskites,
CsNdNb$_2$O$_7$ and RbNdNb$_2$O$_7$,
have been examined in dense polycrystalline samples,
and polarization hysteresis loops that substantiate ferroelectricity have been observed at room temperature.
The theoretical mechanism for the spontaneous polarization, ``hybrid improper ferroelectric mechanism,''
induced by a combination of two types of non-polar octahedral rotations,
is confirmed in these two phases.
Our samples show remanent polarizations of 2--3~$\mu$C/cm$^2$,
which are much larger than those obtained in polycrystalline samples with the hybrid improper ferroelectricity reported so far.
A dielectric constant in CsNdNb$_2$O$_7$ exhibits an anomaly at 625 K,
corresponding to the ferroelectric transition, as previously revealed by x-ray and neutron diffractometry.
No dielectric anomaly is observed for RbNdNb$_2$O$_7$ throughout the temperature range studied here ($\leq 773~\rm K$),
which is consistent with the previous diffractometry showing the persistence of polar $I2cm$ symmetry up to 790~K.

\end{abstract}



\maketitle

\section{Introduction}

Ferroelectric materials have been intensely studied because of fundamental physics involving underlying mechanism behind non-polar to polar structural phase transitions
as well as various technological applications such as non-volatile memory devices and piezoelectric sensors.
Recently, the development of novel ferroelectrics has been accelerated 
since Benedek and Fennie rationalized a novel design principle of ferroelectric layered perovskite oxides utilizing rotations of oxygen-coordinated octahedra.~\cite{Benedek2011PRL}
In their work, they theoretically revisited the origin of polar distortions in $n$~=~2 Ruddlesden-Popper (RP) layered perovskites Ca$_3$Ti$_2$O$_7$ and Ca$_3$Mn$_2$O$_7$,
where $n$ is the number of perovskite layers per block.
They proposed that in these layered perovskites, a spontaneous polarization is induced symmetrically and energetically by a combination of two non-polar rotation modes of oxygen octahedra
via trilinear coupling.
This is called hybrid improper ferroelectricity.~\cite{Bousquet2008Nature,Benedek2011PRL,Rondinelli2012AM,Benedek2012JSSC,Benedek2015DT}
Such crystal chemistry-based guidelines for lifting inversion symmetry have led to the discovery of a bunch of novel acentric, polar and ferroelectric RP oxides.
For example, ferroelectric switching was experimentally demonstrated for $n$~=~2 RP phases (Ca,Sr)$_3$Ti$_2$O$_7$,~\cite{Oh2015NatMater} Ca$_3$(Ti,Mn)$_2$O$_7$,~\cite{Liu2015APL,Liu2018APL}
Sr$_3$Zr$_2$O$_7$,~\cite{Yoshida2018AFM} and (Sr,Ba,Ca)$_3$Sn$_2$O$_7$.~\cite{Wang2016AM,Lu2019APL,Chen2020APL}
In addition, (Ca,Sr,Tb)$_3$(Ti,Fe)$_2$O$_7$ was reported to be polar and potentially ferroelectric.~\cite{Pitcher2015Science}
It was also reported that acentricity is induced by oxygen octahedral rotations in $A$-site ordered $n$ = 1 RP titanates
$AR$TiO$_4$ ($A$ = H, Li, Na, K; $R$ = rare earth).~\cite{Akamatsu2014PRL,SenGupta2016AEM,SenGupta2017CM,Akamatsu2019PRM}

Hybrid improper ferroelectricity is also possible for
$n$~=~2 Dion-Jacobson (DJ) phases,~\cite{Benedek2014IC,Sim2014PRB}
which is another type of layered perovskites consisting of primitive stacking of adjacent perovskite blocks,
in contrast to body-centered stacking in RP phases. 
In spite of the difference in stacking fashions,
ferroelectricity induced by oxygen octahedral rotations was also theoretically predicted for the $n$~=~2 DJ phases $ARB_2$O$_7$ ($A$ = Rb, Cs, $R$ = La, Nd, $B$ = Nb, Ta),~\cite{Benedek2014IC,Sim2014PRB}
and indeed, these oxides were proven to be polar by means of synchrotron x-ray diffraction, neutron diffraction, and optical second harmonic generation measurements.~\cite{Strayer2016AFM,Zhu2017CM,Zhu2020CM}
However, ferroelectric switching has not been demonstrated yet for these $n$~=~2 DJ phases.
On the other hand, ferroelectricity has been reported for Bi-containing DJ phases such as RbBiNb$_2$O$_7$ and CsBiNb$_2$O$_7$,~\cite{Li2012CM,Chen2015JMC}
in which 6s$^2$ stereochemically active lone-pair electrons in Bi$^{3+}$ could play a crucial role in the ferroelectricity.
It was also reported that ferroelectricity is induced by interfacial coupling effects in tailor-made nanosheets composed of $n$~=~2 DJ phases.~\cite{Osada2018DT}

Here, we focus on Bi-free $n$~=~2 DJ phases CsNdNb$_2$O$_7$ and RbNdNb$_2$O$_7$.
Their ferroelectricity is induced almost purely by octahedral rotations as predicted theoretically.~\cite{Benedek2014IC,Sim2014PRB}
Detailed structural analyses for these compounds were recently performed by Zhu $et$ $al$.~\cite{Zhu2017CM,Zhu2020CM}
The crystal structure of CsNdNb$_2$O$_7$ belongs to a polar space group $P2_1am$ with the $a$ and $c$ axes pointing to the directions along the spontaneous polarization and perpendicular to the layer, respectively,~\cite{Snedden2003JSSC,Zhu2017CM}
and a combination of synchrotron x-ray and neutron diffraction recently clarified temperature-induced phase transitions
from $P2_1am$ to $C2/m$ at 625~K to $P4/mmm$ at 800~K.~\cite{Zhu2020CM}
On the other hand, RbNdNb$_2$O$_7$ crystalizes into a structure with polar $I2cm$ space group symmetry, and shows phase transitions from $I2cm$ to $Cmca$ at 790~K to $I4/mcm$ at 865~K.~\cite{Zhu2020CM}
The structural analyses suggested that these compounds are polar and potentially ferroelectric at room temperature, but their ferroelectricity is not reported yet.

In this work, we prepared highly dense and insulating pellets of the $n$~=~2 DJ niobates CsNdNb$_2$O$_7$ and RbNdNb$_2$O$_7$ by a conventional solid-state reaction method,
and demonstrated their ferroelectric switching.
The polycrystalline pellets were characterized by synchrotron x-ray diffraction.
Electric polarization-field ($P$-$E$) curve measurements were performed at room temperature to demonstrate their ferroelectricity.
The ceramic pellets of CsNdNb$_2$O$_7$ and RbNdNb$_2$O$_7$ show the clear $P$-$E$ hysteresis loops with the remanent polarizations
much larger than those observed for polycrystalline samples of other hybrid improper ferroelectrics such as Ca$_3$Ti$_2$O$_7$ and Sr$_3$Zr$_2$O$_7$.
Impedance spectroscopy revealed that the ceramics pellets are highly insulating in contrast to the ceramic pellets of $n$~=~2 DJ niobates reported in previous studies,
allowing for the clear observation of ferroelectric switching of CsNdNb$_2$O$_7$ and RbNdNb$_2$O$_7$.
The polar to non-polar phase transition at 625~K in CsNdNb$_2$O$_7$ was detected by the measurement of temperature-dependent dielectric constants,
while no phase transition was observed for RbNdNb$_2$O$_7$ throughout the measured temperature range ($\leq 773~\rm K$).
These results are in good agreement with the recent structural analyses.~\cite{Zhu2020CM}

\section{Experimental procedures}\label{Exp}

The powder and polycrystalline pellet samples of CsNdNb$_2$O$_7$ and RbNdNb$_2$O$_7$ were prepared via a conventional solid-state reaction method.
Reagent-grade Cs$_2$CO$_3$ (99.9\%; Kojundo Chemical Laboratory Co.~Ltd), Rb$_2$CO$_3$ (99\%; Kojundo Chemical Laboratory Co.~Ltd),
Nd$_2$O$_3$ (99.9\%; Nippon Yttrium Co.~Ltd), and Nb$_2$O$_5$ (99.9\%; Sigma-Aldrich Co.~Ltd) powders were used as starting materials.
The Nd$_2$O$_3$ and Nb$_2$O$_5$ powders were heated at 900~$^\circ$C for 12~h
to eliminate water and carbon dioxide adsorbed on the powders prior to weighing.
A 50~mol\% excess amount of Cs$_2$CO$_3$ and Rb$_2$CO$_3$ were added to a mixture of the starting materials
to compensate for the loss due to the alkaline evaporation during heating.
The mixtures were ground for several tens of minutes with an agate mortar using ethanol as dispersion solvent,
followed by drying at 353~K for 3~h.
The mixture pellets were obtained by uniaxial pressing at 40~MPa  for 1~min and subsequent cold isotropic pressing at 200~MPa for 5~min.
The pellets were put on a Pt plate in an alumina crucible with a lid, and calcined at 1123~K for 12~h.
The calcined pellets were ground, thoroughly mixed, pelletized again,
and then sintered at 1273~K for 48~h.
The phase purity of the samples was confirmed using a laboratory x-ray diffraction (XRD) equipment (D8 ADVANCE; Bruker AXS GmbH).
Relative density of the pellets was determined by the Archimedes' method.

High-resolution synchrotron XRD patterns were taken at room temperature with a Debye-Scherrer camera with MYTHEN solid-state detectors
installed at the BL02B2 beamline of SPring-8.~\cite{Kawaguchi2017RSI}
The ground powder samples were housed in Lindeman capillary tubes with an inner diameter of 0.1~mm.
The capillary tubes were rotated continuously during the measurements to diminish the effect of preferred orientation.
We used an x-ray beam monochromated at $\lambda \rm = 0.670479~or~0.669111~\AA$.
The wavelength is selected so as to prevent the effect of Nb K-edge absorption (0.6532~\AA) and detect the diffraction originating from the largest lattice spacing ($\simeq~11~\rm \AA$) within the 2$\theta$ angle range for the beamline ($\geq 2^\circ$).
Rietveld refinements~\cite{Rietveld} were performed for the XRD patterns using the \textsc{fullprof} suite.~\cite{Rodriguez1993PhysB} 
Bond valence sums of cations were calculated for the refined structures with bond valence parameters provided in Ref.~\onlinecite{Brown1985AC}.
The microstructure on the fractured cross section of a CsNdNb$_2$O$_7$ pellet was observed by field-emission scanning electron microscopy (SEM) using a Hitachi S-5200 SEM.
A Radiant Precision LCI\hspace{-.1em}I ferroelectric tester and a Trek 609B voltage amplifier were used to record $P$-$E$ curves at room temperature for pellet samples with sputtered Pt film electrodes.
The remanent polarization was measured by a pulsed positive-up-negative-down (PUND) polarization measurement technique,
which allows us to extract the switching components of polarization.
The voltage amplitude was varied from 2 to 6~kV with a frequency of 10~Hz.
Electrical impedance spectra were recorded using a 1260A impedance/gain-phase analyzer (Solartron Analytical) with AC modulation of 100~mV covering frequency range from 10 to 10$^6$~Hz
for pellet samples mounted on a temperature controlling stage (LINKAM).
The \textsc{vesta} program was used to visualize crystal structures.~\cite{Momma2008JAC}

\section{Results and Discussion}

\begin{figure}
 \includegraphics[width=8.6cm]{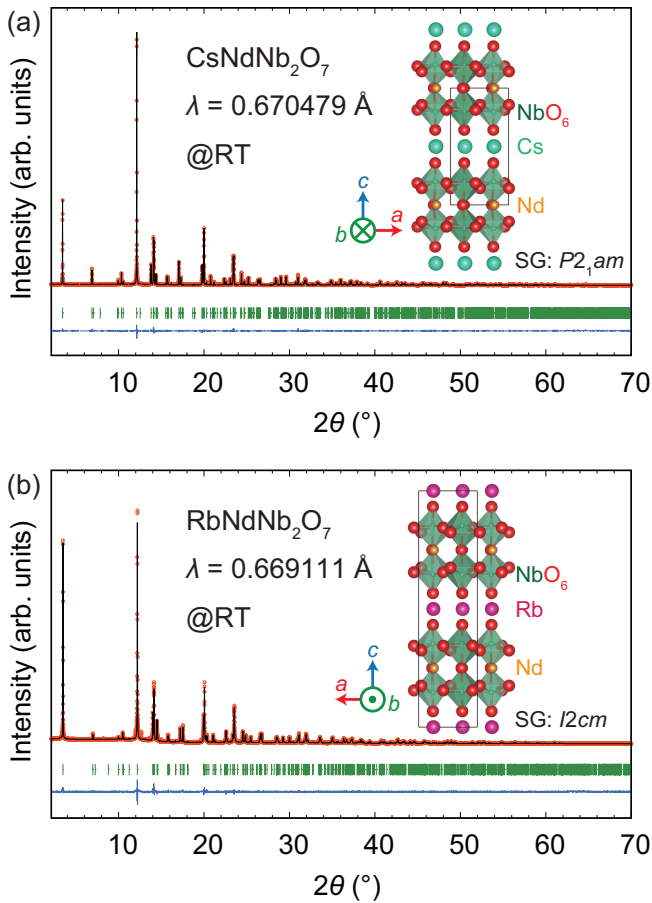}
 \caption{
Synchrotron x-ray diffraction patterns (red open circles) at room temperature for (a) CsNdNb$_2$O$_7$ and (b) RbNdNb$_2$O$_7$
and fitting curves (black lines) obtained by Rietveld refinements
using $P2_{1}am$ and $I2cm$ models for CsNdNb$_2$O$_7$ and RbNdNb$_2$O$_7$, respectively.
The green tics and blue lines represent the positions of Bragg diffractions and the difference between the observed and simulated intensity, respectively.
The insets illustrate the schematics of refined structures.
The solid lines indicate unit cells.
The refined structural parameters are summarized in Tables~\ref{rietveld_Cs} and \ref{rietveld_Rb} for CsNdNb$_2$O$_7$ and RbNdNb$_2$O$_7$, respectively.
}
\label{sxrd}
\end{figure}

Figures~\ref{sxrd}(a) and (b) show the room-temperature synchrotron XRD patterns for the CsNdNb$_2$O$_7$ and RbNdNb$_2$O$_7$ powders, respectively,
as well as fitting curves obtained by Rietveld refinements.
The diffraction peaks for CsNdNb$_2$O$_7$ are indexed by assuming $P2_1am$ symmetry, although a trace amount of NdNbO$_4$ impurity (0.3~wt\%) was present.
The Rietveld refinements were performed for the XRD pattern of CsNdNb$_2$O$_7$ with a $P2_1am$ model~\cite{Snedden2003JSSC} excluding the region of the impurity peaks.
The resultant fit provides small reliability factors:
$R_{\rm wp} = 6.79\%$, $R_{\rm B} = 2.02\%$, and $\chi^2 = 10.44$.
On the other hand, as for RbNdNb$_2$O$_7$,
all the peaks except for the subtle NdNbO$_4$ impurity signals (2.3~wt\%) were well-fitted using an $I2cm$ model~\cite{Zhu2017CM}
with small reliability factors:
$R_{\rm wp} = 11.5\%$, $R_{\rm B} = 3.41\%$, and $\chi^2 = 5.93$.
The refined structural parameters shown in Tables~\ref{rietveld_Cs} and~\ref{rietveld_Rb} are in good agreement with those reported in the previous x-ray and neutron diffraction studies.~\cite{Snedden2003JSSC,Zhu2017CM}
Bond valence sums of the cations calculated for the refined structures are
1.1 for Cs, 2.9 for Nd, and 4.9 for Nb in CsNdNb$_2$O$_7$ and 0.9 for Rb, 3.1 for Nd, and 5.0 for Nb in RbNdNb$_2$O$_7$,
all of which are very close to the formal valences.
The origin of their spontaneous polarizations and the structural difference between the $P2_1am$ and $I2cm$ phases are briefly reviewed in the appendix section~\ref{origin}
in line with the previous theoretical and experimental studies.~\cite{Benedek2014IC,Sim2014PRB,Zhu2017CM}

\begin{table}
\caption{
Structural parameters of CsNdNb$_{2}$O$_{7}$ at room temperature obtained from Rietveld refinement
with a $P2_{1}am$ model against the synchrotron XRD data shown in Fig.~\ref{sxrd}(a).
}
\begin{threeparttable}
\begin{tabular}{lccccc} \toprule
Atom & Site   & $x$        &  $y$       & $z$        & $U_{\rm iso}$ or $U_{\rm eq}$ (\AA$^{2}$) \\
\hline
      Cs\tnote{*}   & 2$b$   & 0.75\tnote{$\dagger$} & 0.7369(3) & $\frac{1}{2}$     & 0.0183(4)   \\
      Nd   & 2$a$   & 0.7591(5) & 0.7473(3)  & 0 & 0.00721(15)    \\
      Nb   & 4$c$   & 0.2422(6) & 0.7505(3)  & 0.20174(4) & 0.00733(14)   \\
      O1   & 4$c$   & 0.238(2) & 0.8008(9) & 0.6420(3) & 0.0053(12)   \\
      O2   & 4$c$   & 0.4325(13) & 0.4436(12) & 0.1867(4) & 0.0120(12)   \\
      O3 & 4$c$   & 0.007(3) & 0.018(2) & 0.1425(4)          & $= U_{\rm iso}$(O2)  \\
      O4 & 2$a$   & 0.2153(19)  & 0.6815(11)  & 0 & $= U_{\rm iso}$(O1)  \\
\hline
\end{tabular}
\begin{tablenotes}\footnotesize
      \item{}Space group; $P2_{1}am$ (No.~26), $Z = 2$. The occupancy parameter is fixed to unity for all atoms.
      $^{\dagger}$Fixed to define an origin of the polar $a$-axis.
      $^{*}$Refined anisotropically.
      Cell parameters: $a$~=~5.470498(6)~\AA, $b$~=~5.449788(7)~\AA, and $c$~=~11.159235(14)~\AA.
      $R_{\mathrm{wp}}$~=~6.79\%, $R_{\mathrm{B}}$~=~2.02\%, and $\chi^{2}$~=~10.44.
\end{tablenotes} 
\end{threeparttable}
\label{rietveld_Cs}
\end{table}

\begin{table}
\caption{
Structural parameters of RbNdNb$_{2}$O$_{7}$ at room temperature obtained from Rietveld refinement
with a $I2cm$ model against synchrotron XRD data shown in Fig.~\ref{sxrd}(b).
}
\begin{threeparttable}
\begin{tabular}{lccccc} \toprule
Atom & Site   & $x$        &  $y$       & $z$        & $U_{\rm iso}$ or $U_{\rm eq}$ (\AA$^{2}$) \\
\hline
      Rb\tnote{*}   & 4$a$   & 0.75\tnote{$\dagger$} &  0 & 0     & 0.0277(18)   \\
      Nd   & 4$b$   & 0.2379(18) & 0.4995(11)  & $\frac{1}{4}$ & 0.0089(3)    \\
      Nb   & 8$c$   & 0.7608(18) & 0.5010(13)  & 0.64664(4) & 0.0100(4)   \\
      O1   & 8$c$   & 0.757(4) & 0.546(2) & 0.4327(2) & 0.0042(15)   \\
      O2   & 8$c$   & 0.581(3) & 0.799(3) & 0.6556(4) & $= U_{\rm iso}$(O1)   \\
      O3  & 8$c$   & 0.957(3) & 0.235(4) & 0.6754(4)          & $= U_{\rm iso}$(O1)  \\
      O4  & 4$b$   & 0.268(6)  & 0.065(3)  & $\frac{1}{4}$ & $= U_{\rm iso}$(O1)  \\
\hline
\end{tabular}
\begin{tablenotes}\footnotesize      
      \item{}Space group; $I2cm$ (No.~46), $Z = 4$. The occupancy parameter is fixed to unity for all atoms.
      $^{\dagger}$Fixed to define an origin of the polar $a$-axis.
      $^{*}$Refined anisotropically.
      Cell parameters: $a$~=~5.44225(2)~\AA, $b$~=~5.42977(2)~\AA, and $c$~=~21.96323(7)~\AA.
      $R_{\mathrm{wp}}$~=~11.5\%, $R_{\mathrm{B}}$~=~3.41\%, and $\chi^{2}$~=~5.93.
\end{tablenotes} 
\end{threeparttable}
\label{rietveld_Rb}
\end{table}

\begin{figure*}
 \includegraphics[width=17cm]{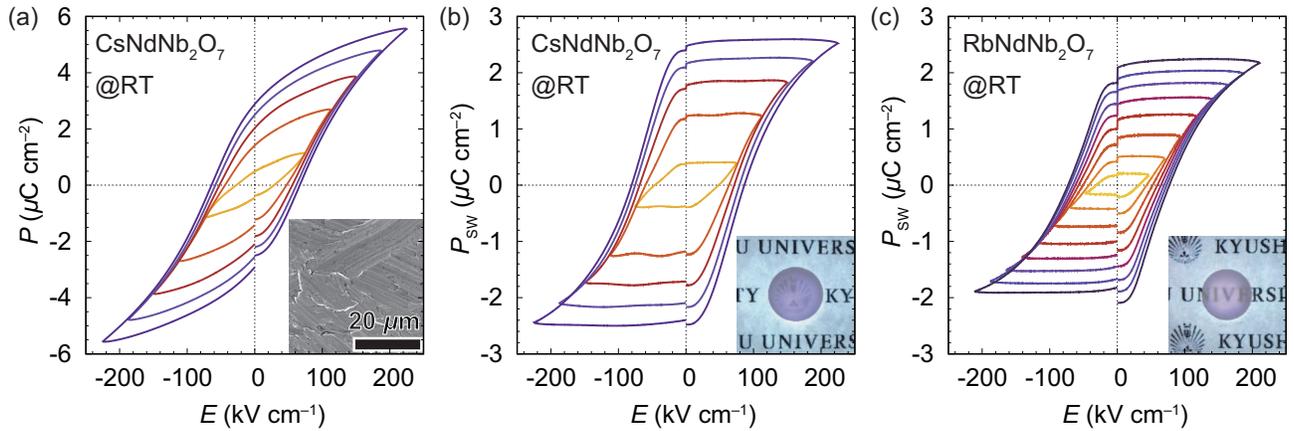}
 \caption{
(a) Polarization ($P$)-field ($E$) curves for CsNdNb$_2$O$_7$ measured at room temperature with varying electric field amplitude,
along with an SEM image of the fractured cross section of the CsNdNb$_2$O$_7$ pellet in the inset.
Switching component of polarization, $P_{\rm sw}$, measured by a PUND method as a function of $E$ for (b) CsNdNb$_2$O$_7$ and (c) RbNdNb$_2$O$_7$,
along with the appearance of the pellets in the insets.
}
\label{p-e}
\end{figure*}

The inset of Fig.~\ref{p-e}(a) shows the cross-sectional SEM image of the CsNdNb$_2$O$_7$ pellet,
confirming the high densification degree.
The microstructure with parallel stripes is similar to that observed by piezo-response force microscopy for the CsBiNb$_2$O$_7$ pellet
prepared by a spark plasma sintering method.~\cite{Chen2015JMC}
The insets of Figs.~\ref{p-e}(b) and (c) show the appearance of the pellet samples of CsNdNb$_2$O$_7$ and RbNdNb$_2$O$_7$, respectively.
Both pellets are translucent because of their small thickness (200--300~$\mu$m) and high relative density (98--99\%).
The light violet color of the pellets stems from optical absorption corresponding to Nd 4f-4f transitions.

Figure~\ref{p-e}(a) displays the $P$-$E$ curves for the CsNdNb$_2$O$_7$ pellet at room temperature.
The $P$-$E$ curves clearly exhibit hysteresis behavior characteristic of ferroelectric materials,
free from piscis-shaped loops due to leakage current,
which are often observed for $n$~=~2 DJ compounds.~\cite{Chen2015JMC}
Only a few reports have indisputably demonstrated ferroelectricity for polar $n$~=~2 DJ phases.
This is partially ascribed to the fact that
this class of samples often show electric leakage and breakdown,~\cite{Goff2009CM,Strayer2016AFM}
hampering the observation of obvious $P$-$E$ hysteresis loops to evidence ferroelectricity.
In a sharp contrast, the present pellets are highly insulating, as referred to the impedance results hereinafter.
The switching component of polarization ($P_{\rm sw}$) measured by a PUND method is plotted against $E$ for CsNdNb$_2$O$_7$ and RbNdNb$_2$O$_7$ in Figs.~\ref{p-e}(b) and (c), respectively.
The observed hysteresis curves verify the ferroelectricity of CsNdNb$_2$O$_7$ and RbNdNb$_2$O$_7$.
The remanent polarization ($P_{\rm r}$), which is a value of $P_{\rm sw}$ at zero field, is increased with an increase in the electric field amplitude,
and $P_{\rm r}$ is not saturated yet even for the maximum field amplitude for both the samples.
The higher fields were not applicable due to electric breakdown.
The maximum $P_{\rm r}$ values were recorded as 2.5 and 2.1~$\mu$C/cm$^2$ for CsNdNb$_2$O$_7$ and RbNdNb$_2$O$_7$, respectively,
which are much smaller than the theoretically calculated polarization, $P_{\rm calc}$, of 29 and 30~$\mu$C/cm$^2$ for CsNdNb$_2$O$_7$ and RbNdNb$_2$O$_7$, respectively.~\cite{Benedek2014IC}
This is because the samples are polycrystalline and the polarizations are not saturated yet due to their high coercive electric fields. 
Compared to the present compounds, higher remanent polarizations were reported for Bi-containing $n$~=~2 DJ phases CsBiNb$_2$O$_7$ and RbBiNb$_2$O$_7$ ($\sim$$10~\mu\rm C/cm^2$),~\cite{Li2012CM,Chen2015JMC}
partly because they have large theoretical polarizations owing to the 6s$^2$ lone-pair electrons in Bi$^{3+}$
($P_{\rm calc}~=~36{\rm -}48~\mu{\rm C/cm^2}$ for CsBiNb$_2$O$_7$ and RbBiNb$_2$O$_7$)~\cite{Fennie2006APL,Benedek2014IC,Sim2014PRB}
and/or because a non-switching component of polarizations originating from electric leakage might be included in the reported remanent polarizations.

It should be noted that the remanent polarizations of CsNdNb$_2$O$_7$ and RbNdNb$_2$O$_7$ are several times larger than those observed for the polycrystalline samples of other hybrid improper ferroelectrics
such as RP layered perovskites Ca$_3$Ti$_2$O$_7$, Sr$_3$Zr$_2$O$_7$, and Sr$_3$Sn$_2$O$_7$,
as summarized in Table~\ref{polarization}.
Now we discuss the reasons behind this,
although it is hard to strictly compare their remanent polarization characteristics because the polarization was recorded with different electric field amplitude ($180{\rm -}320~\rm kV/cm$).
One might think that the larger remanent polarizations of the two DJ phases simply reflect the fact
that $P_{\rm calc}$, i.e., the theoretically predicted potential polarization, is higher for the DJ phases than for the RP phases (See Table~\ref{polarization}).
However, the ratios $P_{\rm r}$/$P_{\rm calc}$ are also larger for the DJ phases than the RP phases,
which excludes the magnitude of potential polarizations from one of the main reasons.
Another reason may be that the coercive electric fields, $E_{\rm c}$, are significantly lower for the DJ phases than for the RP phases,
as shown in Table~\ref{polarization}.
In general, coercive fields depend on various factors such as the Curie temperatures (see Table~\ref{polarization}), the energy barriers against polarization switching and domain wall motions,
and the orientation of polarization axes of each grain in polycrystalline pellets.
To get an insight into the larger polarizations observed for the DJ phases, further experimental studies with single crystals~\cite{Oh2015NatMater} and theoretical investigations on switching pathways and domain wall motions~\cite{Nowadnick2016PRB} are required.

\begin{table}
\caption{
Data regarding electric polarization-field curve measurements for the ceramic pellets of hybrid improper ferroelectrics with layered perovskite structures.
The observed remanent polarization, $P_{\rm r}$ ($\mu$C/cm$^2$), and the theoretically calculated polarization, $P_{\rm calc}$ ($\mu$C/cm$^2$), 
the $P_{\rm r}$/$P_{\rm calc}$ ratios, the coercive field, $E_{\rm c}$ (kV/cm), and the electric field amplitude used in the measurements, $E_{\rm meas}$ (kV/cm),
are given as well as Curie temperature, $T_{\rm C}$ (K).
The numbers in the brackets indicate the references, from which the data are cited.
}
\begin{threeparttable}
\begin{tabular}{lcccccc} \toprule
Compound & $P_{\rm r}$ & $P_{\rm calc}$ &  $P_{\rm r}$/$P_{\rm calc}$ & $E_{\rm c}$ & $E_{\rm meas}$ & $T_{\rm C}$ \\
\hline
      CsNdNb$_2$O$_7$ & 2.6  & 29 [\onlinecite{Benedek2014IC}] &  0.086 & 90 & 230 & 625 [\onlinecite{Zhu2020CM}]\\
      RbNdNb$_2$O$_7$ & 2.1 & 30 [\onlinecite{Benedek2014IC}] & 0.070 & 80 & 210 & 790 [\onlinecite{Zhu2020CM}]\\
      Ca$_3$Ti$_2$O$_7$ & 0.6 [\onlinecite{Liu2015APL}]  & 20 [\onlinecite{Benedek2011PRL}] & 0.03 & 120 & 240 & 1100 [\onlinecite{Liu2015APL}]\\
      Sr$_3$Zr$_2$O$_7$ & 0.3 [\onlinecite{Yoshida2018AFM}] & 7.2 [\onlinecite{Mulder2013AFM}] & 0.04 & 150 & 180 & 700 [\onlinecite{Yoshida2018AFM}]\\
      Sr$_3$Sn$_2$O$_7$ & 0.24 [\onlinecite{Wang2016AM}] & 3.9 [\onlinecite{Mulder2013AFM}] & 0.062 & 200 & 320 & 490 [\onlinecite{Yoshida2018JACS}] \\
  \hline
\end{tabular}
\end{threeparttable}
\label{polarization}
\end{table}

\begin{figure*}
 \includegraphics[width=18cm]{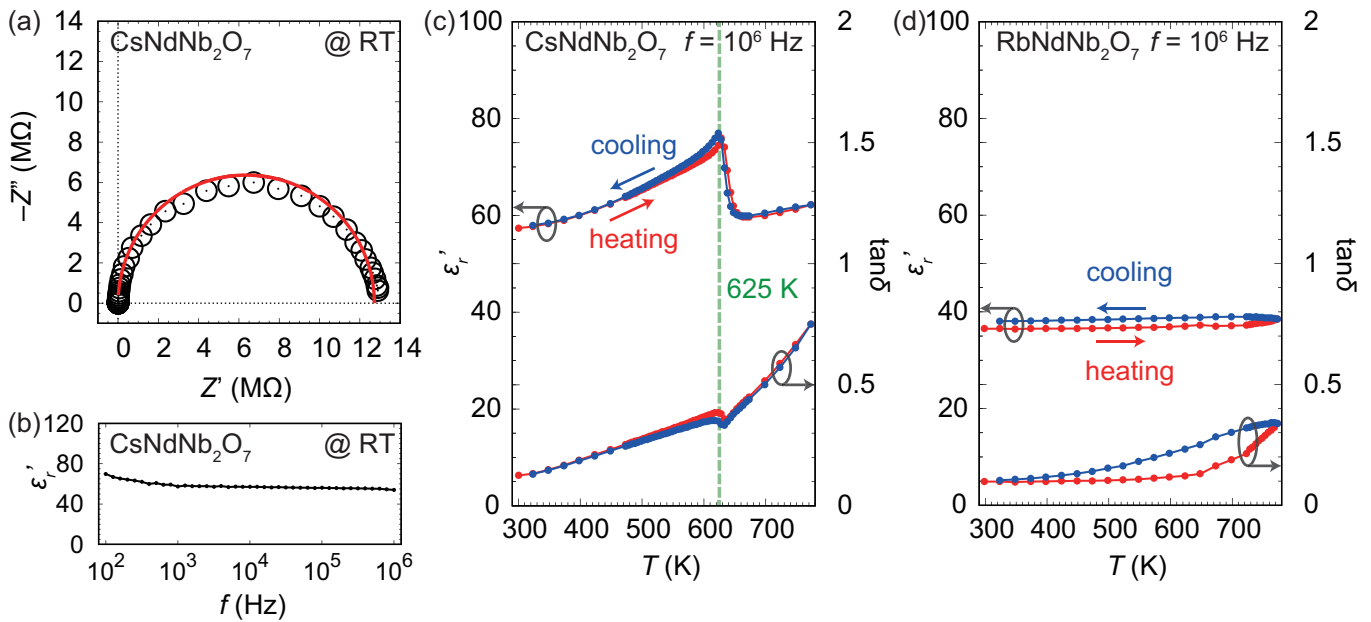}
 \caption{
(a) Nyquist plot at room temperature for the CsNdNb$_2$O$_7$ pellet.
The red semicircle presents a fitting curve obtained using an equivalent circuit model composed of a single, parallel RC element.
(b) Frequency dependence of the real part of relative dielectric constant, $\epsilon^\prime_r$, for CsNdNb$_2$O$_7$.
Temperature dependence of $\epsilon^\prime_r$ and the dielectric loss, tan$\delta$,
for the (c) CsNdNb$_2$O$_7$ and (d) RbNdNb$_2$O$_7$ pellets measured at 10$^6$~Hz.
}
\label{epsilon}
\end{figure*}

Impedance spectroscopy was performed to investigate the electrical properties and dielectric constants.
Figure~\ref{epsilon}(a) presents a Nyquist plot for the CsNdNb$_2$O$_7$ pellet at room temperature.
The observed semicircle was analyzed using an equivalent circuit model comprising a single, parallel RC element,
resulting in a fairly good fit with a resistance of $\sim$13~M$\Omega$.
The insulating nature of the pellet sample is confirmed by the high resistance.
Such a highly insulating property allows for the clear observation of the polarization switching shown in Fig.~\ref{p-e}. 
The electrical conductivity of the pellet is 9.4$\times$10$^{-9}$~S/cm,
which is much lower than that reported for $n$~=~2 DJ phase CsBiNb$_2$O$_7$
($\sim$10$^{-3}$~S/cm)
due to significant proton conductivity.~\cite{Goff2009CM}
The real and imaginary parts of relative dielectric constants, $\epsilon^\prime_r$ and $\epsilon^{\prime\prime}_r$, respectively, were estimated with the equivalent circuit model.
Figure~\ref{epsilon}(b) plots $\epsilon^\prime_r$ at room temperature as a function of frequency.
There is little frequency dependence of $\epsilon^\prime_r$ ($\simeq$~55) in a frequency range from $10^3$ and $10^6$~Hz,
while a weak upturn is observed below $10^3$~Hz.
Figure~\ref{epsilon}(c) shows the temperature dependence of $\epsilon^\prime_r$ and the dielectric loss, tan$\delta$ (= $\epsilon^{\prime\prime}_r$/$\epsilon^\prime_r$),
for the CsNdNb$_2$O$_7$ pellet measured at $f$ = $10^6$~Hz.
A dielectric anomaly is observed at 625~K both on heating and cooling with a small thermal hysteresis.
This observation of anomaly coincides with the recent structural analysis revealing that
CsNdNb$_2$O$_7$ exhibits a phase transition from polar $P2_1am$ to non-polar $C2/m$ phases at 625~K on heating.~\cite{Zhu2020CM}
The dielectric loss is low at room temperature (tan$\delta$~$\simeq$~0.1).
 The overall increase in tan$\delta$ with increasing temperature indicates that the electrical leakage becomes significant at higher temperatures. 
 On the other hand, $\epsilon^\prime_r$ of the RbNdNb$_2$O$_7$ pellet is almost constant ($\epsilon^\prime_r = 36{\rm -}40$)
 without any dielectric anomaly throughout the temperature range studied here ($\leq$ 773~K),
 in accordance with the recent report that RbNdNb$_2$O$_7$ keeps polar $I2cm$ symmetry up to 790~K.~\cite{Zhu2020CM}
 The RbNdNb$_2$O$_7$ pellet also shows a low dielectric loss at room temperature (tan$\delta$~$\simeq$~0.1),
 revealing its highly insulating nature.
 The value of $\epsilon^\prime_r$ at room temperature in our samples are comparable to those in other $n$~=~2 DJ niobates ($\epsilon^\prime_r = 10{\rm -}60$).~\cite{Li2012CM,Dixon2017Crystals,Chen2015JMC,Goff2009CM}

 \section{Conclusion}\label{conclusion}
 
 Through the observation of polarization switching by electric field,
 we clearly demonstrate the ferroelectricity of $n$~=~2 DJ layered perovskites CsNdNb$_2$O$_7$ and RbNdNb$_2$O$_7$,
 which have been regarded as potentially ferroelectric phases according to the previous structural analyses so far.
 The remanent polarizations of the polycrystalline pellets (2--3~$\mu$C/cm$^2$) are much larger than those of other hybrid improper ferroelectrics with layered perovskite structures.
 The clear $P$-$E$ hysteresis loops were obtained thanks to the high electrical resistivity, as confirmed by the impedance spectroscopy.
 The temperature-dependent relative dielectric constants exhibit the anomaly at 625~K for CsNdNb$_2$O$_7$ reflecting a ferroelectric phase transition,
 while no dielectric anomaly was detected for RbNdNb$_2$O$_7$ up to the highest temperature studied here (773~K).

\begin{acknowledgments}

The synchrotron radiation experiments were performed at the BL02B2 of SPring-8
with the approval of the Japan Synchrotron Radiation Research Institute (JASRI)
(Proposal No.~2018A1152 and No.~2018B1227).
This research was supported by
Japan Society of the Promotion of Science (JSPS) KAKENHI Grants No.~JP16H06440, No.~JP17K19172, and No.~JP18H01892,
and by Murata Science Foundation.

\end{acknowledgments}

\appendix

 \section{The origin of spontaneous polarizations and the structural difference between the $P2_1am$ and $I2cm$ phases}\label{origin}
 
Here let us review the origin of spontaneous polarizations in the $P2_1am$ and $I2cm$ phases and their structural difference.~\cite{Benedek2014IC,Sim2014PRB,Zhu2017CM}
The $P2_1am$ symmetry is established by a direct sum of irreducible representations (irreps) M$_5^-$ and M$_2^+$ for the aristotype, undistorted $P4/mmm$ phase.
The atomic displacement transforming like M$_5^-$ corresponds to oxygen octahedral tilting represented by $a^-a^-c^0$ in Glazer notation,~\cite{Glazer1972AC}
and the M$_2^+$ distortion corresponds to rotation represented by $a^0a^0c^+$;
therefore, the $P2_1am$ phase shows an $a^-a^-c^+$-type of octahedral rotation.
The rotation and tilting distortions with a wavenumber vector at the M point ($\frac{1}{2}$,$\frac{1}{2}$,0) lead to a cell expansion to a $\sqrt{2}a_t\times\sqrt{2}a_t\times c_t$ metric,
where $a_t$ and $c_t$ are lattice constants of the aristotype $P4/mmm$ phase.
The $P2_1am$ structure also encompasses a polar distortion transforming like an irrep $\Gamma_5^-$,
which generates $P$ along the $a$ axis.
The polar $\Gamma_5^-$ distortion is induced by a combination of the M$_2^+$ and M$_5^-$ distortions in terms of both symmetry and energy lowering;
a product of order parameters of the M$_5^-$, M$_2^+$, and $\Gamma_5^-$ distortions is invariant under the symmetry operations of $P4/mmm$,
so the Landau energy expansion includes a trilinear coupling term:
\begin{eqnarray}
F_{\rm tri} \propto \eta ({\rm M}_5^-) \eta ({\rm M}_2^+) P,
\label{eq:ftri}
\end{eqnarray}
where $\eta({\rm M}_5^-)$ and $\eta({\rm M}_2^+)$ are order parameters of the M$_5^-$ and M$_2^+$ distortions, respectively,
and the polarization $P$ is proportional to the order parameter of the $\Gamma_5^-$ distortion.

In the $I2cm$ structure, each perovskite slab shows an $a^-a^-c^+$-type of octahedral rotation,
but the sense of rotation and tilting is opposite to that in the adjacent slabs.
Therefore, the rotation pattern is represented as $a^-a^-c^+$/--($a^-a^-c^+$)
using an extended version of Glazer notation defined in Ref.~\onlinecite{Zhu2017CM},
where rotation patterns of the lower and higher slabs are written before and behind a slash,
and the opposite sense of rotation and tilting is denoted by a prefactor minus sign.
The $I2cm$ symmetry is established by a direct sum of irreps A$_5^-$ and A$_2^+$ for the $P4/mmm$ phase.~\cite{Zhu2017CM}
The atomic displacement transforming like the irrep A$_5^-$ corresponds to an $a^-a^-c^0$/--$a^-$--$a^-c^0$-type of tilting,
and the A$_2^+$ distortion corresponds to an $a^0a^0c^+$/$a^0a^0$--$c^+$-type of rotation.
The distortion modes at the A point ($\frac{1}{2}$,$\frac{1}{2}$,$\frac{1}{2}$) result in a cell expansion to a $\sqrt{2}a_t\times\sqrt{2}a_t\times 2c_t$ metric.
Hence, the conventional unit cell of the $I2cm$ phase is twice as long as that of the $P2_1am$ phase [see the insets of Figs.~\ref{sxrd}(a) and (b)].
A polar $\Gamma_5^-$ distortion is induced by the following trilinear coupling involving the A$_5^-$ and A$_2^+$ modes,
the form of which is similar to that for the $P2_1am$ phase:
\begin{eqnarray}
F_{\rm tri} \propto \eta({\rm A}_5^-)\eta({\rm A}_2^+)P,
\label{eq:ftri}
\end{eqnarray}
where $\eta({\rm A}_5^-)$ and $\eta({\rm A}_2^+)$ are order parameters of the A$_5^-$ and A$_2^+$ distortions, respectively.


\begin{thebibliography}{38}
\expandafter\ifx\csname natexlab\endcsname\relax\def\natexlab#1{#1}\fi
\expandafter\ifx\csname bibnamefont\endcsname\relax
  \def\bibnamefont#1{#1}\fi
\expandafter\ifx\csname bibfnamefont\endcsname\relax
  \def\bibfnamefont#1{#1}\fi
\expandafter\ifx\csname citenamefont\endcsname\relax
  \def\citenamefont#1{#1}\fi
\expandafter\ifx\csname url\endcsname\relax
  \def\url#1{\texttt{#1}}\fi
\expandafter\ifx\csname urlprefix\endcsname\relax\def\urlprefix{URL }\fi
\providecommand{\bibinfo}[2]{#2}
\providecommand{\eprint}[2][]{\url{#2}}

\bibitem[{\citenamefont{Benedek and Fennie}(2011)}]{Benedek2011PRL}
\bibinfo{author}{\bibfnamefont{N.~A.} \bibnamefont{Benedek}} \bibnamefont{and}
  \bibinfo{author}{\bibfnamefont{C.~J.} \bibnamefont{Fennie}},
  \bibinfo{journal}{Phys. Rev. Lett.} \textbf{\bibinfo{volume}{106}},
  \bibinfo{pages}{107204} (\bibinfo{year}{2011}).

\bibitem[{\citenamefont{Bousquet et~al.}(2008)\citenamefont{Bousquet, Dawber,
  Stucki, Lichtensteiger, Hermet, Gariglio, Triscone, and
  Ghosez}}]{Bousquet2008Nature}
\bibinfo{author}{\bibfnamefont{E.}~\bibnamefont{Bousquet}},
  \bibinfo{author}{\bibfnamefont{M.}~\bibnamefont{Dawber}},
  \bibinfo{author}{\bibfnamefont{N.}~\bibnamefont{Stucki}},
  \bibinfo{author}{\bibfnamefont{C.}~\bibnamefont{Lichtensteiger}},
  \bibinfo{author}{\bibfnamefont{P.}~\bibnamefont{Hermet}},
  \bibinfo{author}{\bibfnamefont{S.}~\bibnamefont{Gariglio}},
  \bibinfo{author}{\bibfnamefont{J.-M.} \bibnamefont{Triscone}},
  \bibnamefont{and} \bibinfo{author}{\bibfnamefont{P.}~\bibnamefont{Ghosez}},
  \bibinfo{journal}{Nature} \textbf{\bibinfo{volume}{452}},
  \bibinfo{pages}{732} (\bibinfo{year}{2008}).

\bibitem[{\citenamefont{Rondinelli and Fennie}(2012)}]{Rondinelli2012AM}
\bibinfo{author}{\bibfnamefont{J.~M.} \bibnamefont{Rondinelli}}
  \bibnamefont{and} \bibinfo{author}{\bibfnamefont{C.~J.}
  \bibnamefont{Fennie}}, \bibinfo{journal}{Adv. Mater.}
  \textbf{\bibinfo{volume}{24}}, \bibinfo{pages}{1961} (\bibinfo{year}{2012}).

\bibitem[{\citenamefont{Benedek et~al.}(2012)\citenamefont{Benedek, Mulder, and
  Fennie}}]{Benedek2012JSSC}
\bibinfo{author}{\bibfnamefont{N.~A.} \bibnamefont{Benedek}},
  \bibinfo{author}{\bibfnamefont{A.~T.} \bibnamefont{Mulder}},
  \bibnamefont{and} \bibinfo{author}{\bibfnamefont{C.~J.}
  \bibnamefont{Fennie}}, \bibinfo{journal}{J. Solid State Chem.}
  \textbf{\bibinfo{volume}{195}}, \bibinfo{pages}{11 } (\bibinfo{year}{2012}).

\bibitem[{\citenamefont{Benedek et~al.}(2015)\citenamefont{Benedek, Rondinelli,
  Djani, Ghosez, and Lightfoot}}]{Benedek2015DT}
\bibinfo{author}{\bibfnamefont{N.~A.} \bibnamefont{Benedek}},
  \bibinfo{author}{\bibfnamefont{J.~M.} \bibnamefont{Rondinelli}},
  \bibinfo{author}{\bibfnamefont{H.}~\bibnamefont{Djani}},
  \bibinfo{author}{\bibfnamefont{P.}~\bibnamefont{Ghosez}}, \bibnamefont{and}
  \bibinfo{author}{\bibfnamefont{P.}~\bibnamefont{Lightfoot}},
  \bibinfo{journal}{Dalton Trans.} \textbf{\bibinfo{volume}{44}},
  \bibinfo{pages}{10543} (\bibinfo{year}{2015}).

\bibitem[{\citenamefont{Oh et~al.}(2015)\citenamefont{Oh, Luo, Huang, Wang, and
  Cheong}}]{Oh2015NatMater}
\bibinfo{author}{\bibfnamefont{Y.~S.} \bibnamefont{Oh}},
  \bibinfo{author}{\bibfnamefont{X.}~\bibnamefont{Luo}},
  \bibinfo{author}{\bibfnamefont{F.-T.} \bibnamefont{Huang}},
  \bibinfo{author}{\bibfnamefont{Y.}~\bibnamefont{Wang}}, \bibnamefont{and}
  \bibinfo{author}{\bibfnamefont{S.-W.} \bibnamefont{Cheong}},
  \bibinfo{journal}{Nature Mater.} \textbf{\bibinfo{volume}{14}},
  \bibinfo{pages}{407} (\bibinfo{year}{2015}).

\bibitem[{\citenamefont{Liu et~al.}(2015)\citenamefont{Liu, Wu, Shi, Zhao,
  Zhou, Qiu, Zhang, and Chen}}]{Liu2015APL}
\bibinfo{author}{\bibfnamefont{X.~Q.} \bibnamefont{Liu}},
  \bibinfo{author}{\bibfnamefont{J.~W.} \bibnamefont{Wu}},
  \bibinfo{author}{\bibfnamefont{X.~X.} \bibnamefont{Shi}},
  \bibinfo{author}{\bibfnamefont{H.~J.} \bibnamefont{Zhao}},
  \bibinfo{author}{\bibfnamefont{H.~Y.} \bibnamefont{Zhou}},
  \bibinfo{author}{\bibfnamefont{R.~H.} \bibnamefont{Qiu}},
  \bibinfo{author}{\bibfnamefont{W.~Q.} \bibnamefont{Zhang}}, \bibnamefont{and}
  \bibinfo{author}{\bibfnamefont{X.~M.} \bibnamefont{Chen}},
  \bibinfo{journal}{Appl. Phys. Lett.} \textbf{\bibinfo{volume}{106}},
  \bibinfo{pages}{202903} (\bibinfo{year}{2015}).

\bibitem[{\citenamefont{Liu et~al.}(2018)\citenamefont{Liu, Zhang, Lin, Lin,
  Yang, Li, Wang, Li, Yan, Wang et~al.}}]{Liu2018APL}
\bibinfo{author}{\bibfnamefont{M.}~\bibnamefont{Liu}},
  \bibinfo{author}{\bibfnamefont{Y.}~\bibnamefont{Zhang}},
  \bibinfo{author}{\bibfnamefont{L.-F.} \bibnamefont{Lin}},
  \bibinfo{author}{\bibfnamefont{L.}~\bibnamefont{Lin}},
  \bibinfo{author}{\bibfnamefont{S.}~\bibnamefont{Yang}},
  \bibinfo{author}{\bibfnamefont{X.}~\bibnamefont{Li}},
  \bibinfo{author}{\bibfnamefont{Y.}~\bibnamefont{Wang}},
  \bibinfo{author}{\bibfnamefont{S.}~\bibnamefont{Li}},
  \bibinfo{author}{\bibfnamefont{Z.}~\bibnamefont{Yan}},
  \bibinfo{author}{\bibfnamefont{X.}~\bibnamefont{Wang}}, \bibnamefont{et~al.},
  \bibinfo{journal}{Appl. Phys. Lett.} \textbf{\bibinfo{volume}{113}},
  \bibinfo{pages}{022902} (\bibinfo{year}{2018}).

\bibitem[{\citenamefont{Yoshida
  et~al.}(2018{\natexlab{a}})\citenamefont{Yoshida, Fujita, Akamatsu,
  Hernandez, {Sen Gupta}, Brown, Padmanabhan, Gibbs, Kuge, Tsuji
  et~al.}}]{Yoshida2018AFM}
\bibinfo{author}{\bibfnamefont{S.}~\bibnamefont{Yoshida}},
  \bibinfo{author}{\bibfnamefont{K.}~\bibnamefont{Fujita}},
  \bibinfo{author}{\bibfnamefont{H.}~\bibnamefont{Akamatsu}},
  \bibinfo{author}{\bibfnamefont{O.}~\bibnamefont{Hernandez}},
  \bibinfo{author}{\bibfnamefont{A.}~\bibnamefont{{Sen Gupta}}},
  \bibinfo{author}{\bibfnamefont{F.~G.} \bibnamefont{Brown}},
  \bibinfo{author}{\bibfnamefont{H.}~\bibnamefont{Padmanabhan}},
  \bibinfo{author}{\bibfnamefont{A.~S.} \bibnamefont{Gibbs}},
  \bibinfo{author}{\bibfnamefont{T.}~\bibnamefont{Kuge}},
  \bibinfo{author}{\bibfnamefont{R.}~\bibnamefont{Tsuji}},
  \bibnamefont{et~al.}, \bibinfo{journal}{Adv. Funct. Mater.}
  \textbf{\bibinfo{volume}{28}}, \bibinfo{pages}{1801856}
  (\bibinfo{year}{2018}{\natexlab{a}}).

\bibitem[{\citenamefont{Wang et~al.}(2017)\citenamefont{Wang, Huang, Luo, Gao,
  and Cheong}}]{Wang2016AM}
\bibinfo{author}{\bibfnamefont{Y.}~\bibnamefont{Wang}},
  \bibinfo{author}{\bibfnamefont{F.-T.} \bibnamefont{Huang}},
  \bibinfo{author}{\bibfnamefont{X.}~\bibnamefont{Luo}},
  \bibinfo{author}{\bibfnamefont{B.}~\bibnamefont{Gao}}, \bibnamefont{and}
  \bibinfo{author}{\bibfnamefont{S.-W.} \bibnamefont{Cheong}},
  \bibinfo{journal}{Adv. Mater.} \textbf{\bibinfo{volume}{29}},
  \bibinfo{pages}{1601288} (\bibinfo{year}{2017}).

\bibitem[{\citenamefont{Lu et~al.}(2019)\citenamefont{Lu, Liu, Ma, Fu, Yuan,
  Wu, and Chen}}]{Lu2019APL}
\bibinfo{author}{\bibfnamefont{J.~J.} \bibnamefont{Lu}},
  \bibinfo{author}{\bibfnamefont{X.~Q.} \bibnamefont{Liu}},
  \bibinfo{author}{\bibfnamefont{X.}~\bibnamefont{Ma}},
  \bibinfo{author}{\bibfnamefont{M.~S.} \bibnamefont{Fu}},
  \bibinfo{author}{\bibfnamefont{A.}~\bibnamefont{Yuan}},
  \bibinfo{author}{\bibfnamefont{Y.~J.} \bibnamefont{Wu}}, \bibnamefont{and}
  \bibinfo{author}{\bibfnamefont{X.~M.} \bibnamefont{Chen}},
  \bibinfo{journal}{Journal of Applied Physics} \textbf{\bibinfo{volume}{125}},
  \bibinfo{pages}{044101} (\bibinfo{year}{2019}).

\bibitem[{\citenamefont{Chen et~al.}(2020)\citenamefont{Chen, Sun, Liu, Zhu,
  Tian, and Chen}}]{Chen2020APL}
\bibinfo{author}{\bibfnamefont{B.~H.} \bibnamefont{Chen}},
  \bibinfo{author}{\bibfnamefont{T.~L.} \bibnamefont{Sun}},
  \bibinfo{author}{\bibfnamefont{X.~Q.} \bibnamefont{Liu}},
  \bibinfo{author}{\bibfnamefont{X.~L.} \bibnamefont{Zhu}},
  \bibinfo{author}{\bibfnamefont{H.}~\bibnamefont{Tian}}, \bibnamefont{and}
  \bibinfo{author}{\bibfnamefont{X.~M.} \bibnamefont{Chen}},
  \bibinfo{journal}{Applied Physics Letters} \textbf{\bibinfo{volume}{116}},
  \bibinfo{pages}{042903} (\bibinfo{year}{2020}).

\bibitem[{\citenamefont{Pitcher et~al.}(2015)\citenamefont{Pitcher, Mandal,
  Dyer, Alaria, Borisov, Niu, Claridge, and Rosseinsky}}]{Pitcher2015Science}
\bibinfo{author}{\bibfnamefont{M.~J.} \bibnamefont{Pitcher}},
  \bibinfo{author}{\bibfnamefont{P.}~\bibnamefont{Mandal}},
  \bibinfo{author}{\bibfnamefont{M.~S.} \bibnamefont{Dyer}},
  \bibinfo{author}{\bibfnamefont{J.}~\bibnamefont{Alaria}},
  \bibinfo{author}{\bibfnamefont{P.}~\bibnamefont{Borisov}},
  \bibinfo{author}{\bibfnamefont{H.}~\bibnamefont{Niu}},
  \bibinfo{author}{\bibfnamefont{J.~B.} \bibnamefont{Claridge}},
  \bibnamefont{and} \bibinfo{author}{\bibfnamefont{M.~J.}
  \bibnamefont{Rosseinsky}}, \bibinfo{journal}{Science}
  \textbf{\bibinfo{volume}{347}}, \bibinfo{pages}{420} (\bibinfo{year}{2015}).

\bibitem[{\citenamefont{Akamatsu et~al.}(2014)\citenamefont{Akamatsu, Fujita,
  Kuge, Sen~Gupta, Togo, Lei, Xue, Stone, Rondinelli, Chen
  et~al.}}]{Akamatsu2014PRL}
\bibinfo{author}{\bibfnamefont{H.}~\bibnamefont{Akamatsu}},
  \bibinfo{author}{\bibfnamefont{K.}~\bibnamefont{Fujita}},
  \bibinfo{author}{\bibfnamefont{T.}~\bibnamefont{Kuge}},
  \bibinfo{author}{\bibfnamefont{A.}~\bibnamefont{Sen~Gupta}},
  \bibinfo{author}{\bibfnamefont{A.}~\bibnamefont{Togo}},
  \bibinfo{author}{\bibfnamefont{S.}~\bibnamefont{Lei}},
  \bibinfo{author}{\bibfnamefont{F.}~\bibnamefont{Xue}},
  \bibinfo{author}{\bibfnamefont{G.}~\bibnamefont{Stone}},
  \bibinfo{author}{\bibfnamefont{J.~M.} \bibnamefont{Rondinelli}},
  \bibinfo{author}{\bibfnamefont{L.-Q.} \bibnamefont{Chen}},
  \bibnamefont{et~al.}, \bibinfo{journal}{Phys. Rev. Lett.}
  \textbf{\bibinfo{volume}{112}}, \bibinfo{pages}{187602}
  (\bibinfo{year}{2014}).

\bibitem[{\citenamefont{Gupta et~al.}(2016)\citenamefont{Gupta, Akamatsu,
  Strayer, Lei, Kuge, Fujita, dela Cruz, Togo, Tanaka, Tanaka
  et~al.}}]{SenGupta2016AEM}
\bibinfo{author}{\bibfnamefont{A.~S.} \bibnamefont{Gupta}},
  \bibinfo{author}{\bibfnamefont{H.}~\bibnamefont{Akamatsu}},
  \bibinfo{author}{\bibfnamefont{M.~E.} \bibnamefont{Strayer}},
  \bibinfo{author}{\bibfnamefont{S.}~\bibnamefont{Lei}},
  \bibinfo{author}{\bibfnamefont{T.}~\bibnamefont{Kuge}},
  \bibinfo{author}{\bibfnamefont{K.}~\bibnamefont{Fujita}},
  \bibinfo{author}{\bibfnamefont{C.}~\bibnamefont{dela Cruz}},
  \bibinfo{author}{\bibfnamefont{A.}~\bibnamefont{Togo}},
  \bibinfo{author}{\bibfnamefont{I.}~\bibnamefont{Tanaka}},
  \bibinfo{author}{\bibfnamefont{K.}~\bibnamefont{Tanaka}},
  \bibnamefont{et~al.}, \bibinfo{journal}{Adv. Electron. Mater.}
  \textbf{\bibinfo{volume}{2}}, \bibinfo{pages}{1500196}
  (\bibinfo{year}{2016}).

\bibitem[{\citenamefont{{Sen Gupta} et~al.}(2017)\citenamefont{{Sen Gupta},
  Akamatsu, Brown, Nguyen, Strayer, Lapidus, Yoshida, Fujita, Tanaka, Tanaka
  et~al.}}]{SenGupta2017CM}
\bibinfo{author}{\bibfnamefont{A.}~\bibnamefont{{Sen Gupta}}},
  \bibinfo{author}{\bibfnamefont{H.}~\bibnamefont{Akamatsu}},
  \bibinfo{author}{\bibfnamefont{F.~G.} \bibnamefont{Brown}},
  \bibinfo{author}{\bibfnamefont{M.~A.~T.} \bibnamefont{Nguyen}},
  \bibinfo{author}{\bibfnamefont{M.~E.} \bibnamefont{Strayer}},
  \bibinfo{author}{\bibfnamefont{S.}~\bibnamefont{Lapidus}},
  \bibinfo{author}{\bibfnamefont{S.}~\bibnamefont{Yoshida}},
  \bibinfo{author}{\bibfnamefont{K.}~\bibnamefont{Fujita}},
  \bibinfo{author}{\bibfnamefont{K.}~\bibnamefont{Tanaka}},
  \bibinfo{author}{\bibfnamefont{I.}~\bibnamefont{Tanaka}},
  \bibnamefont{et~al.}, \bibinfo{journal}{Chem. Mater.}
  \textbf{\bibinfo{volume}{29}}, \bibinfo{pages}{656} (\bibinfo{year}{2017}).

\bibitem[{\citenamefont{Akamatsu et~al.}(2019)\citenamefont{Akamatsu, Fujita,
  Kuge, Gupta, Rondinelli, Tanaka, Tanaka, and Gopalan}}]{Akamatsu2019PRM}
\bibinfo{author}{\bibfnamefont{H.}~\bibnamefont{Akamatsu}},
  \bibinfo{author}{\bibfnamefont{K.}~\bibnamefont{Fujita}},
  \bibinfo{author}{\bibfnamefont{T.}~\bibnamefont{Kuge}},
  \bibinfo{author}{\bibfnamefont{A.~S.} \bibnamefont{Gupta}},
  \bibinfo{author}{\bibfnamefont{J.~M.} \bibnamefont{Rondinelli}},
  \bibinfo{author}{\bibfnamefont{I.}~\bibnamefont{Tanaka}},
  \bibinfo{author}{\bibfnamefont{K.}~\bibnamefont{Tanaka}}, \bibnamefont{and}
  \bibinfo{author}{\bibfnamefont{V.}~\bibnamefont{Gopalan}},
  \bibinfo{journal}{Phys. Rev. Mater.} \textbf{\bibinfo{volume}{3}},
  \bibinfo{pages}{065001} (\bibinfo{year}{2019}).

\bibitem[{\citenamefont{Benedek}(2014)}]{Benedek2014IC}
\bibinfo{author}{\bibfnamefont{N.~A.} \bibnamefont{Benedek}},
  \bibinfo{journal}{Inorg. Chem.} \textbf{\bibinfo{volume}{53}},
  \bibinfo{pages}{3769} (\bibinfo{year}{2014}).

\bibitem[{\citenamefont{Sim and Kim}(2014)}]{Sim2014PRB}
\bibinfo{author}{\bibfnamefont{H.}~\bibnamefont{Sim}} \bibnamefont{and}
  \bibinfo{author}{\bibfnamefont{B.~G.} \bibnamefont{Kim}},
  \bibinfo{journal}{Phys. Rev. B} \textbf{\bibinfo{volume}{89}},
  \bibinfo{pages}{144114} (\bibinfo{year}{2014}).

\bibitem[{\citenamefont{Strayer et~al.}(2016)\citenamefont{Strayer, Gupta,
  Akamatsu, Lei, Benedek, Gopalan, and Mallouk}}]{Strayer2016AFM}
\bibinfo{author}{\bibfnamefont{M.~E.} \bibnamefont{Strayer}},
  \bibinfo{author}{\bibfnamefont{A.~S.} \bibnamefont{Gupta}},
  \bibinfo{author}{\bibfnamefont{H.}~\bibnamefont{Akamatsu}},
  \bibinfo{author}{\bibfnamefont{S.}~\bibnamefont{Lei}},
  \bibinfo{author}{\bibfnamefont{N.~A.} \bibnamefont{Benedek}},
  \bibinfo{author}{\bibfnamefont{V.}~\bibnamefont{Gopalan}}, \bibnamefont{and}
  \bibinfo{author}{\bibfnamefont{T.~E.} \bibnamefont{Mallouk}},
  \bibinfo{journal}{Adv. Funct. Mater.} \textbf{\bibinfo{volume}{26}},
  \bibinfo{pages}{1930} (\bibinfo{year}{2016}).

\bibitem[{\citenamefont{Zhu et~al.}(2017)\citenamefont{Zhu, Cohen, Gibbs,
  Zhang, Halasyamani, Hayward, and Benedek}}]{Zhu2017CM}
\bibinfo{author}{\bibfnamefont{T.}~\bibnamefont{Zhu}},
  \bibinfo{author}{\bibfnamefont{T.}~\bibnamefont{Cohen}},
  \bibinfo{author}{\bibfnamefont{A.~S.} \bibnamefont{Gibbs}},
  \bibinfo{author}{\bibfnamefont{W.}~\bibnamefont{Zhang}},
  \bibinfo{author}{\bibfnamefont{P.~S.} \bibnamefont{Halasyamani}},
  \bibinfo{author}{\bibfnamefont{M.~A.} \bibnamefont{Hayward}},
  \bibnamefont{and} \bibinfo{author}{\bibfnamefont{N.~A.}
  \bibnamefont{Benedek}}, \bibinfo{journal}{Chem. Mater.}
  \textbf{\bibinfo{volume}{29}}, \bibinfo{pages}{9489} (\bibinfo{year}{2017}).

\bibitem[{\citenamefont{Zhu et~al.}(2020)\citenamefont{Zhu, Gibbs, Benedek, and
  Hayward}}]{Zhu2020CM}
\bibinfo{author}{\bibfnamefont{T.}~\bibnamefont{Zhu}},
  \bibinfo{author}{\bibfnamefont{A.~S.} \bibnamefont{Gibbs}},
  \bibinfo{author}{\bibfnamefont{N.~A.} \bibnamefont{Benedek}},
  \bibnamefont{and} \bibinfo{author}{\bibfnamefont{M.~A.}
  \bibnamefont{Hayward}}, \bibinfo{journal}{Chem. Mater.}
  \textbf{\bibinfo{volume}{32}}, \bibinfo{pages}{4340} (\bibinfo{year}{2020}).

\bibitem[{\citenamefont{Li et~al.}(2012)\citenamefont{Li, Osada, Ozawa, and
  Sasaki}}]{Li2012CM}
\bibinfo{author}{\bibfnamefont{B.~W.} \bibnamefont{Li}},
  \bibinfo{author}{\bibfnamefont{M.}~\bibnamefont{Osada}},
  \bibinfo{author}{\bibfnamefont{T.~C.} \bibnamefont{Ozawa}}, \bibnamefont{and}
  \bibinfo{author}{\bibfnamefont{T.}~\bibnamefont{Sasaki}},
  \bibinfo{journal}{Chem. Mater.} \textbf{\bibinfo{volume}{24}},
  \bibinfo{pages}{3111} (\bibinfo{year}{2012}).

\bibitem[{\citenamefont{Chen et~al.}(2015)\citenamefont{Chen, Ning, Lepadatu,
  Cain, Yan, and Reece}}]{Chen2015JMC}
\bibinfo{author}{\bibfnamefont{C.}~\bibnamefont{Chen}},
  \bibinfo{author}{\bibfnamefont{H.}~\bibnamefont{Ning}},
  \bibinfo{author}{\bibfnamefont{S.}~\bibnamefont{Lepadatu}},
  \bibinfo{author}{\bibfnamefont{M.}~\bibnamefont{Cain}},
  \bibinfo{author}{\bibfnamefont{H.}~\bibnamefont{Yan}}, \bibnamefont{and}
  \bibinfo{author}{\bibfnamefont{M.~J.} \bibnamefont{Reece}},
  \bibinfo{journal}{J. Mater. Chem. C} \textbf{\bibinfo{volume}{3}},
  \bibinfo{pages}{19} (\bibinfo{year}{2015}).

\bibitem[{\citenamefont{Osada and Sasaki}(2018)}]{Osada2018DT}
\bibinfo{author}{\bibfnamefont{M.}~\bibnamefont{Osada}} \bibnamefont{and}
  \bibinfo{author}{\bibfnamefont{T.}~\bibnamefont{Sasaki}},
  \bibinfo{journal}{Dalton Trans.} \textbf{\bibinfo{volume}{47}},
  \bibinfo{pages}{2841} (\bibinfo{year}{2018}).

\bibitem[{\citenamefont{Snedden et~al.}(2003)\citenamefont{Snedden, Knight, and
  Lightfoot}}]{Snedden2003JSSC}
\bibinfo{author}{\bibfnamefont{A.}~\bibnamefont{Snedden}},
  \bibinfo{author}{\bibfnamefont{K.~S.} \bibnamefont{Knight}},
  \bibnamefont{and}
  \bibinfo{author}{\bibfnamefont{P.}~\bibnamefont{Lightfoot}},
  \bibinfo{journal}{J. Solid State Chem.} \textbf{\bibinfo{volume}{173}},
  \bibinfo{pages}{309 } (\bibinfo{year}{2003}).

\bibitem[{\citenamefont{Kawaguchi et~al.}(2017)\citenamefont{Kawaguchi,
  Takemoto, Osaka, Nishibori, Moriyoshi, Kubota, Kuroiwa, and
  Sugimoto}}]{Kawaguchi2017RSI}
\bibinfo{author}{\bibfnamefont{S.}~\bibnamefont{Kawaguchi}},
  \bibinfo{author}{\bibfnamefont{M.}~\bibnamefont{Takemoto}},
  \bibinfo{author}{\bibfnamefont{K.}~\bibnamefont{Osaka}},
  \bibinfo{author}{\bibfnamefont{E.}~\bibnamefont{Nishibori}},
  \bibinfo{author}{\bibfnamefont{C.}~\bibnamefont{Moriyoshi}},
  \bibinfo{author}{\bibfnamefont{Y.}~\bibnamefont{Kubota}},
  \bibinfo{author}{\bibfnamefont{Y.}~\bibnamefont{Kuroiwa}}, \bibnamefont{and}
  \bibinfo{author}{\bibfnamefont{K.}~\bibnamefont{Sugimoto}},
  \bibinfo{journal}{Rev. Sci. Instrum.} \textbf{\bibinfo{volume}{88}},
  \bibinfo{pages}{085111} (\bibinfo{year}{2017}).

\bibitem[{\citenamefont{Rietveld}(1969)}]{Rietveld}
\bibinfo{author}{\bibfnamefont{H.~M.} \bibnamefont{Rietveld}},
  \bibinfo{journal}{Journal of Applied Crystallography}
  \textbf{\bibinfo{volume}{2}}, \bibinfo{pages}{65} (\bibinfo{year}{1969}).

\bibitem[{\citenamefont{Rodriguez-Carvajal}(1993)}]{Rodriguez1993PhysB}
\bibinfo{author}{\bibfnamefont{J.}~\bibnamefont{Rodriguez-Carvajal}},
  \bibinfo{journal}{Physica B} \textbf{\bibinfo{volume}{192}},
  \bibinfo{pages}{55} (\bibinfo{year}{1993}).

\bibitem[{\citenamefont{Brown and Altermatt}(1985)}]{Brown1985AC}
\bibinfo{author}{\bibfnamefont{I.~D.} \bibnamefont{Brown}} \bibnamefont{and}
  \bibinfo{author}{\bibfnamefont{D.}~\bibnamefont{Altermatt}},
  \bibinfo{journal}{Acta Crystallogr. B} \textbf{\bibinfo{volume}{41}},
  \bibinfo{pages}{244} (\bibinfo{year}{1985}).

\bibitem[{\citenamefont{Momma and Izumi}(2008)}]{Momma2008JAC}
\bibinfo{author}{\bibfnamefont{K.}~\bibnamefont{Momma}} \bibnamefont{and}
  \bibinfo{author}{\bibfnamefont{F.}~\bibnamefont{Izumi}}, \bibinfo{journal}{J.
  Appl. Crystallogr.} \textbf{\bibinfo{volume}{41}}, \bibinfo{pages}{653}
  (\bibinfo{year}{2008}).

\bibitem[{\citenamefont{Goff et~al.}(2009)\citenamefont{Goff, Keeble, Thomas,
  Ritter, Morrison, and Lightfoot}}]{Goff2009CM}
\bibinfo{author}{\bibfnamefont{R.~J.} \bibnamefont{Goff}},
  \bibinfo{author}{\bibfnamefont{D.}~\bibnamefont{Keeble}},
  \bibinfo{author}{\bibfnamefont{P.~A.} \bibnamefont{Thomas}},
  \bibinfo{author}{\bibfnamefont{C.}~\bibnamefont{Ritter}},
  \bibinfo{author}{\bibfnamefont{F.~D.} \bibnamefont{Morrison}},
  \bibnamefont{and}
  \bibinfo{author}{\bibfnamefont{P.}~\bibnamefont{Lightfoot}},
  \bibinfo{journal}{Chem. Mater.} \textbf{\bibinfo{volume}{21}},
  \bibinfo{pages}{1296} (\bibinfo{year}{2009}).

\bibitem[{\citenamefont{Fennie and Rabe}(2006)}]{Fennie2006APL}
\bibinfo{author}{\bibfnamefont{C.~J.} \bibnamefont{Fennie}} \bibnamefont{and}
  \bibinfo{author}{\bibfnamefont{K.~M.} \bibnamefont{Rabe}},
  \bibinfo{journal}{Appl. Phys. Lett.} \textbf{\bibinfo{volume}{88}},
  \bibinfo{pages}{262902} (\bibinfo{year}{2006}).

\bibitem[{\citenamefont{Nowadnick and Fennie}(2016)}]{Nowadnick2016PRB}
\bibinfo{author}{\bibfnamefont{E.~A.} \bibnamefont{Nowadnick}}
  \bibnamefont{and} \bibinfo{author}{\bibfnamefont{C.~J.}
  \bibnamefont{Fennie}}, \bibinfo{journal}{Physical Review B}
  \textbf{\bibinfo{volume}{94}}, \bibinfo{pages}{104105}
  (\bibinfo{year}{2016}).

\bibitem[{\citenamefont{Mulder et~al.}(2013)\citenamefont{Mulder, Benedek,
  Rondinelli, and Fennie}}]{Mulder2013AFM}
\bibinfo{author}{\bibfnamefont{A.~T.} \bibnamefont{Mulder}},
  \bibinfo{author}{\bibfnamefont{N.~A.} \bibnamefont{Benedek}},
  \bibinfo{author}{\bibfnamefont{J.~M.} \bibnamefont{Rondinelli}},
  \bibnamefont{and} \bibinfo{author}{\bibfnamefont{C.~J.}
  \bibnamefont{Fennie}}, \bibinfo{journal}{Adv. Funct. Mater.}
  \textbf{\bibinfo{volume}{23}}, \bibinfo{pages}{4810} (\bibinfo{year}{2013}).

\bibitem[{\citenamefont{Yoshida
  et~al.}(2018{\natexlab{b}})\citenamefont{Yoshida, Akamatsu, Tsuji, Hernandez,
  Padmanabhan, {Sen Gupta}, Gibbs, Mibu, Murai, Rondinelli
  et~al.}}]{Yoshida2018JACS}
\bibinfo{author}{\bibfnamefont{S.}~\bibnamefont{Yoshida}},
  \bibinfo{author}{\bibfnamefont{H.}~\bibnamefont{Akamatsu}},
  \bibinfo{author}{\bibfnamefont{R.}~\bibnamefont{Tsuji}},
  \bibinfo{author}{\bibfnamefont{O.}~\bibnamefont{Hernandez}},
  \bibinfo{author}{\bibfnamefont{H.}~\bibnamefont{Padmanabhan}},
  \bibinfo{author}{\bibfnamefont{A.}~\bibnamefont{{Sen Gupta}}},
  \bibinfo{author}{\bibfnamefont{A.~S.} \bibnamefont{Gibbs}},
  \bibinfo{author}{\bibfnamefont{K.}~\bibnamefont{Mibu}},
  \bibinfo{author}{\bibfnamefont{S.}~\bibnamefont{Murai}},
  \bibinfo{author}{\bibfnamefont{J.~M.} \bibnamefont{Rondinelli}},
  \bibnamefont{et~al.}, \bibinfo{journal}{J. Am. Chem. Soc.}
  \textbf{\bibinfo{volume}{140}}, \bibinfo{pages}{15690}
  (\bibinfo{year}{2018}{\natexlab{b}}).

\bibitem[{\citenamefont{Dixon et~al.}(2017)\citenamefont{Dixon, McNulty,
  Knight, Gibbs, and Lightfoot}}]{Dixon2017Crystals}
\bibinfo{author}{\bibfnamefont{C.}~\bibnamefont{Dixon}},
  \bibinfo{author}{\bibfnamefont{J.}~\bibnamefont{McNulty}},
  \bibinfo{author}{\bibfnamefont{K.}~\bibnamefont{Knight}},
  \bibinfo{author}{\bibfnamefont{A.}~\bibnamefont{Gibbs}}, \bibnamefont{and}
  \bibinfo{author}{\bibfnamefont{P.}~\bibnamefont{Lightfoot}},
  \bibinfo{journal}{Crystals} \textbf{\bibinfo{volume}{7}},
  \bibinfo{pages}{135} (\bibinfo{year}{2017}).

\bibitem[{\citenamefont{Glazer}(1972)}]{Glazer1972AC}
\bibinfo{author}{\bibfnamefont{A.~M.} \bibnamefont{Glazer}},
  \bibinfo{journal}{Acta Crystallogr. B} \textbf{\bibinfo{volume}{28}},
  \bibinfo{pages}{3384} (\bibinfo{year}{1972}).

\end{thebibliography}

\end{document}